\documentclass[preprint,12pt, a4paper]{elsarticle}

\usepackage{amssymb}
\usepackage{pifont}
\newcommand{\xmark}{\ding{55}}%
\newcommand{\cmark}{\ding{51}}%
\usepackage{hyperref}
\usepackage{comment}
\usepackage{listings}
\usepackage{array}
\usepackage{tcolorbox}
\usepackage{blindtext}  
\usepackage{xcolor}
\definecolor{maroon}{cmyk}{0, 0.87, 0.68, 0.32}
\definecolor{halfgray}{gray}{0.55}
\definecolor{ipython_frame}{RGB}{207, 207, 207}
\definecolor{ipython_bg}{RGB}{247, 247, 247}
\definecolor{ipython_red}{RGB}{186, 33, 33}
\definecolor{ipython_green}{RGB}{0, 128, 0}
\definecolor{ipython_cyan}{RGB}{64, 128, 128}
\definecolor{ipython_purple}{RGB}{170, 34, 255}

\usepackage{listings}
\lstdefinelanguage{iPython}{
    morekeywords={access,and,break,class,continue,def,del,elif,else,except,exec,finally,for,from,global,if,import,in,is,lambda,not,or,pass,print,raise,return,try,while},%
    morekeywords=[2]{abs,all,any,basestring,bin,bool,bytearray,callable,chr,classmethod,cmp,compile,complex,delattr,dict,dir,divmod,enumerate,eval,execfile,file,filter,float,format,frozenset,getattr,globals,hasattr,hash,help,hex,id,input,int,isinstance,issubclass,iter,len,list,locals,long,map,max,memoryview,min,next,object,oct,open,ord,pow,property,range,raw_input,reduce,reload,repr,reversed,round,set,setattr,slice,sorted,staticmethod,str,sum,super,tuple,type,unichr,unicode,vars,xrange,zip,apply,buffer,coerce,intern},%
    sensitive=true,%
    morecomment=[l]\#,%
    morestring=[b]',%
    morestring=[b]",%
    morestring=[s]{'''}{'''},
    morestring=[s]{"""}{"""},
    morestring=[s]{r'}{'},
    morestring=[s]{r"}{"},%
    morestring=[s]{r'''}{'''},%
    morestring=[s]{r"""}{"""},%
    morestring=[s]{u'}{'},
    morestring=[s]{u"}{"},%
    morestring=[s]{u'''}{'''},%
    morestring=[s]{u"""}{"""},%
    literate=
    {á}{{\'a}}1 {é}{{\'e}}1 {í}{{\'i}}1 {ó}{{\'o}}1 {ú}{{\'u}}1
    {Á}{{\'A}}1 {É}{{\'E}}1 {Í}{{\'I}}1 {Ó}{{\'O}}1 {Ú}{{\'U}}1
    {à}{{\`a}}1 {è}{{\`e}}1 {ì}{{\`i}}1 {ò}{{\`o}}1 {ù}{{\`u}}1
    {À}{{\`A}}1 {È}{{\'E}}1 {Ì}{{\`I}}1 {Ò}{{\`O}}1 {Ù}{{\`U}}1
    {ä}{{\"a}}1 {ë}{{\"e}}1 {ï}{{\"i}}1 {ö}{{\"o}}1 {ü}{{\"u}}1
    {Ä}{{\"A}}1 {Ë}{{\"E}}1 {Ï}{{\"I}}1 {Ö}{{\"O}}1 {Ü}{{\"U}}1
    {â}{{\^a}}1 {ê}{{\^e}}1 {î}{{\^i}}1 {ô}{{\^o}}1 {û}{{\^u}}1
    {Â}{{\^A}}1 {Ê}{{\^E}}1 {Î}{{\^I}}1 {Ô}{{\^O}}1 {Û}{{\^U}}1
    {œ}{{\oe}}1 {Œ}{{\OE}}1 {æ}{{\ae}}1 {Æ}{{\AE}}1 {ß}{{\ss}}1
    {ç}{{\c c}}1 {Ç}{{\c C}}1 {ø}{{\o}}1 {å}{{\r a}}1 {Å}{{\r A}}1
    {€}{{\EUR}}1 {£}{{\pounds}}1
    {^}{{{\color{ipython_purple}\^{}}}}1
    {=}{{{\color{ipython_purple}=}}}1
    {+}{{{\color{ipython_purple}+}}}1
    {*}{{{\color{ipython_purple}$^\ast$}}}1
    {/}{{{\color{ipython_purple}/}}}1
    {+=}{{{+=}}}1
    {-=}{{{-=}}}1
    {*=}{{{$^\ast$=}}}1
    {/=}{{{/=}}}1,
    literate=
    *{-}{{{\color{ipython_purple}-}}}1
     {?}{{{\color{ipython_purple}?}}}1,
    identifierstyle=\color{black}\ttfamily,
    commentstyle=\color{ipython_cyan}\ttfamily,
    stringstyle=\color{ipython_red}\ttfamily,
    keepspaces=true,
    showspaces=false,
    showstringspaces=false,
    rulecolor=\color{ipython_frame},
    frame=single,
    frameround={t}{t}{t}{t},
    framexleftmargin=6mm,
    numbers=left,
    numberstyle=\tiny\color{halfgray},
    backgroundcolor=\color{ipython_bg},
    basicstyle=\scriptsize,
    keywordstyle=\color{ipython_green}\ttfamily,
}

\begin{document}

\begin{frontmatter}



\title{musicaiz: A Python Library for Symbolic Music Generation, Analysis and Visualization}


\author{Carlos Hernandez-Olivan\corref{corresponding author}}
\ead{carloshero@unizar.es}
\ead[url]{https://carlosholivan.github.io/}

\author{Jose R. Beltran}
\ead{jrbelbla@unizar.es}

\address{Department of Electronic Engineering and Communications, University of Zaragoza, 50018 Zaragoza, Spain}

\begin{abstract}
In this article, we present musicaiz, an object-oriented library for analyzing, generating and evaluating symbolic music. The submodules of the package allow the user to create symbolic music data from scratch, build algorithms to analyze symbolic music, encode MIDI data as tokens to train deep learning sequence models, modify existing music data and evaluate music generation systems. The evaluation submodule builds on previous work to objectively measure music generation systems and to be able to reproduce the results of music generation models. 
The library is publicly available online. We encourage the community to contribute and provide feedback.

\end{abstract}


\end{frontmatter}

\section{Motivation and significance} \label{sec:introduction}

The field of Music Information Retrieval (MIR) research has grown significantly in the last few decades. In recent years, with the growth of Artificial Intelligence, several models have been proposed in the subfields of MIR that are becoming real-world applications, such as music production. The automation and scalability of new models has become a necessity for music companies such as Sony, Spotify, Apple Music, Pandora or Dolby.

One of the fastest growing subfields of MIR in recent years is music generation. The reason for this growing interest is the emergence of new deep learning models that allow researchers to build better performance models versus previous rule-based approaches \cite{review}, \cite{briot2020deep}.

This has increased the interest in building open source tools and software to help researchers in the preprocessing stages when working, on the one hand, with symbolic music data like \texttt{jSymbolic} \cite{jsymbolic}, \texttt{music21} \cite{music21}, \texttt{Humdrum} \cite{humdrum}, \texttt{mido} or \texttt{pretty\_midi} \cite{pretty_midi} and, on the other hand, with audio samples such as \texttt{librosa} \cite{librosa}. Other libraries have been proposed to deal with symbolic music data for music generation purposes such as \texttt{muspy} \cite{muspy} or libraries to help construct encodings for training sequence models such as \texttt{miditok} \cite{miditok}. 

Each of these softwares works on a particular step of the music generation or analysis flow. The goal of \texttt{musicaiz} is to bring together music theory, music processing, models for music generation and evaluation techniques to provide all the steps necessary to build a music generation system that can be reproducible and scalable. The library is publicly available\footnote{\url{https://github.com/carlosholivan/musicaiz}}. We also provide the documentation and examples\footnote{\url{https://carlosholivan.github.io/musicaiz/index.html}}. The software is tested with \texttt{pytest}\footnote{\url{https://github.com/pytest-dev/pytest}}.

In Table \ref{tab:comparison} we compare the features offered by these packages in comparison with the \texttt{musicaiz} library.

\begin{figure}
        \centering
        \includegraphics[width=0.9\columnwidth]{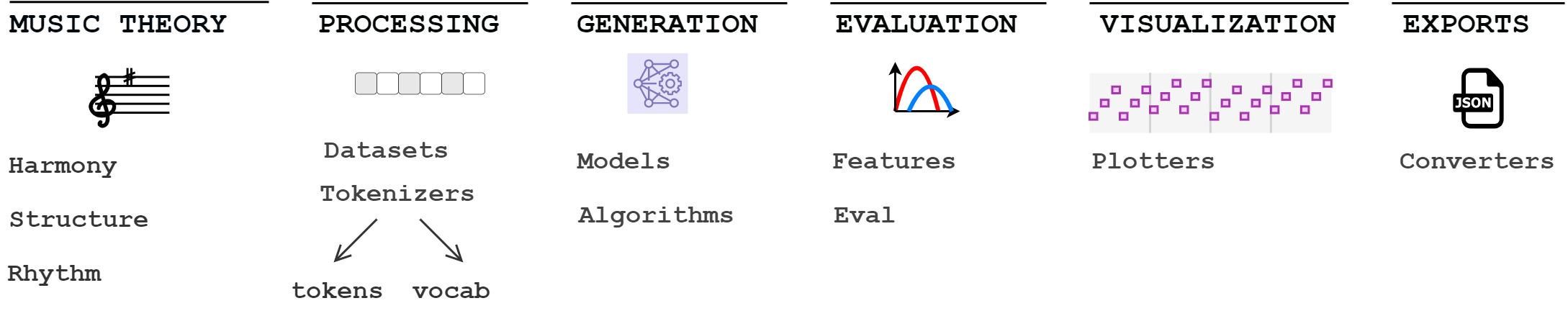}
        \caption{Musicaiz utilities and submodules.}
        \label{fig:scheme_gen}
\end{figure}

\texttt{Musicaiz} library (Fig.\ref{fig:scheme_gen}) is build on \texttt{pretty\_midi} \cite{pretty_midi} and its design principles are aligned with the music theory principles.

\begin{table}[!h]
\begin{tabular}{
    l|
    >{\centering\arraybackslash}p{1.5cm} 
    >{\centering\arraybackslash}p{1.5cm}  
    >{\centering\arraybackslash}p{1.5cm}  
    >{\centering\arraybackslash}p{2cm}  
    >{\centering\arraybackslash}p{1.5cm}  
    >{\centering\arraybackslash}p{1.5cm} 
}
\hline
\textbf{Package} & \textbf{Lang.} & \textbf{Purpose} & \textbf{Music Theory} & \textbf{Tokenizers} & \textbf{Models} & \textbf{Eval.} \\
\hline
jSymbolic & java & analysis & \cmark & \xmark & \xmark & \cmark \\
\hline
music21 & python & general & \cmark & \xmark & \xmark &  \xmark \\
\hline
mido & python & processing & \xmark & \xmark & \xmark & \xmark\\
\hline
pretty\_midi & python & processing & \xmark & \xmark & \xmark & \xmark\\
\hline
Humdrum & awk & analysis & \cmark & \xmark & \xmark & \cmark\\
\hline
Miditok & python & tokenize & \xmark & \cmark & \xmark & \xmark\\
\hline
Muspy & python & generation & \xmark & \xmark & \xmark & \cmark\\
\hline
Musicaiz & python & generation, visual., analysis & \cmark & \cmark & \cmark & \cmark\\
\hline
\end{tabular}
\caption{Frameworks for MIDI and symbolic music generation, analysis and representation.}
\label{tab:comparison} 
\end{table}

\section{Software description}\label{sec:modules}

\subsection{Software architecture}
\texttt{Musicaiz} is written in Python 3. The software is divided into several submodules to suit the user's needs. There are submodules that allow to represent the symbolic format in an OOP approach by following music theory principles such as rhythm, harmony and structure. From these submodules we build a \texttt{features} submodule that allows to extract information from the raw symbolic music data. These features form the basis of the \texttt{eval} submodule that allows us to evaluate music generation systems to help make them reproducible, which is a current need in the music generation subfield \cite{review}.
Since deep learning models often require a significant preprocessing step, especially sequence models such as the Transformer \cite{attention}, we also provide a tokenizer module that allows tokenizing symbolic music data with the Multi-track Music Machine encoding \cite{ens2020mmm}. This submodule can be easily extended to different encodings due to its OOP design. In the \texttt{datasets} submodule, we provide classes that allow processing symbolic music datasets used in the field of music generation.
Furthermore, symbolic music can be represented as a pianoroll and there are no programs that allow to plot these representations with a grid of subdivisions as digital audio workstations (DAWs) do. We provide a plotting submodule that helps to plot symbolic musical pianorolls and save them as HTML files with \texttt{plotly}. This can not only help to visualize the symbolic musical data, but also could be used as a debugging tool.
Finally, the exports submodule converts \texttt{musicaiz} objects to a JSON format so that music information can be sent via REST APIs. This allows building music generation applications with \texttt{musicaiz} on the backend that could be used not only for research purposes, but also in industry.

\subsection{Software Functionalities}
The goal of \texttt{musicaiz} is to provide researchers, practitioners and users with a framework for music generation. We can group the submodules of the software into a high-level abstraction that caters to the software's functionalities (see Fig. \ref{fig:scheme_gen}). These groups are: music theory, processing, generation, evaluation, visualization and export. Although the software is intended to serve as a music generation framework, it could serve more purposes with the configuration of the current submodules.

\subsubsection{Loaders}
The module \texttt{loaders} contains the main class of the library which is called \texttt{Musa}. It allows to import a MIDI file and initialize all its corresponding attributes. The default tempo or bpm value is set to 120 and the resolution or Ticks per Quarter Note (TPQN) is set to 96, however, if a MIDI file is provided, the tempo value will be the MIDI tempo attribute if set in the MIDI metamessages.

\subsubsection{Music Theory}
The main submodules that allow computing and processing symbolic information following the principles of music theory are: rhythm, structure and harmony.

\paragraph{Rhythm} In this module the user can operate with different time signatures and apply different quantizations to data imported from a MIDI file. The classes and methods contained in the \texttt{musicaiz.rhythm} module allow us to group notes into measures, calculate the duration of notes, etc. This module contains two submodules: \texttt{time} and \texttt{quantization}.

\paragraph{Structure}
The structure is another basic submodule of the library. It contains elements referring to musical form or structure such as piece, instrument or notes.
The main elements of this module are are: \texttt{Instrument}, \texttt{Bar} and \texttt{Note}:
\begin{itemize}
    \item \texttt{structure.notes} contains classes that are based on Pearce and Wiggins' work \cite{pearce2006expectation}. It allows the construction of note objects for analysis or symbolic music generation. Notes can contain pitch, time and performance attributes.
    \item \texttt{structure.bars} has a class \texttt{Bar} that helps to define the measures. The purpose of this submodule is to use this class to build music processors and provide more ways to organize symbolic musical data that are aligned with the score representation.
    \item \texttt{structure.instruments} helps to define and group notes and measures within their corresponding instrument.
\end{itemize}

\paragraph{Harmony}
We have built the main structure blocks on the time axis of the music, but the music also works on a vertical or pitch axis. We include a \texttt{harmony} submodule to be able to analyze or generate harmonic data. The main elements of this submodule are: \texttt{Intervals}, \texttt{Chords} and \texttt{Tonalities} and \texttt{Scales}:
\begin{itemize}
    \item \texttt{harmony.intervals} contain definitions of common intervals as defined in music theory \cite{piston_harmony}.
    \item \texttt{harmony.chords} contains the chord definitions. Chords are defined by the root note, its quality, its type or complexity (triads or sevenths) and the list of intervals from the root note containing the chord.
    \item \texttt{harmony.keys}. We define the keys from zero to seventh alterations, sharps and flats. Each tonality and mode (major and minor) has its corresponding scales in major, minor and Greek modes.
\end{itemize}

\subsubsection{Processing} \label{sec:processing}
Token representations allow training deep learning sequence models. In recent years, several works include token representations to train such models in the music generation task. 
We provide the submodules \texttt{tokenizers} and \texttt{datasets} to build token representations and process commonly used open-source music generation datasets.

\paragraph{Tokenizers} In \texttt{musicaiz}, we implemented the tokenization defined in the Multitrack Music Machine model (MMM) \cite{ens2020mmm} which uses the GPT-2 model \cite{gpt2} to generate multi-track music. We can export the tokens in \texttt{.txt} format to be able to use the tokens with other software. We also provide the \texttt{get\_vocabulary} method to save the vocabulary of a given tokenization also as a \texttt{.txt} file.

\paragraph{Datasets} For simplifying the processing of common datasets, we include this submodule that uses the \texttt{tokenizers} submodule to process the following datasets: MAESTRO \cite{hawthorne2018enabling}, Lakh MIDI Dataset \cite{raffel2016learning} and JSB Chorales \cite{jsbchorales}.

\subsubsection{Generation} \label{sec:generation}
For the specific task of symbolic music generation, \texttt{musicaiz} contains two submodules: \texttt{algorithms} and \texttt{models}.

\paragraph{Algorithms} This submodule contains the implementation of a harmonic transposition algorithm. Given the degrees per measure and the scale of a given symbolic music data, it transposes the music to the target degrees and scale. This can serve as a data augmentation technique for training large deep learning models.

\paragraph{Models} This submodule contains another submodule \texttt{transformer\_composers} which contains the implementation of a GPT-based model with its corresponding training and dataloaders modules in Pytorch \cite{NEURIPS2019_9015}. The model can be trained to generate symbolic music. With the submodules mentioned above, we can easily evaluate the results and calculate the preprocessing steps required by the model. The submodule \texttt{models}  can be easily extended to provide more state-of-the-art model implementations for music generation such as the Music Transformer \cite{HuangVUSHSDHDE19}.

\subsubsection{Evaluation} \label{sec:eval}
From the music theory classes and functions, we built a submodule \texttt{features} to extract features from the symbolic music data that we use in the submodule \texttt{eval}.

\paragraph{Features} The submodule \texttt{features} contains implementations of some pitch, rhythm and harmonic features. The goal of this module is to extract features from symbolic music data that can be used to analyze, measure or also build encodings to condition deep learning sequence models. The main elements of this module are: \texttt{Rhythm}, \texttt{Harmony} and \texttt{Pitch} and \texttt{Self-Similarity}:
\begin{itemize}
    \item \texttt{features.rhythm}. We implemented part of the work of Roig et al. \cite{roig2014automatic} to provide a time signature numerator estimation.
    \item \texttt{features.harmony}. Contains a basic algorithm to detect the global key of a MIDI file.
    \item \texttt{features.pitch}. It contains basic pitch characteristics from which the submodule \texttt{eval} is subsequently built, allowing the evaluation of the musical generation models.
    \item \texttt{features.self\_similarity}. Allows to build self-similarity matrices \cite{foote1999visualizing} in the symbolic domain.
\end{itemize} 

\begin{figure*}[h]
    \centering
    \begin{minipage}[b]{0.2\textwidth}
        \centering
        \includegraphics[width=\columnwidth]{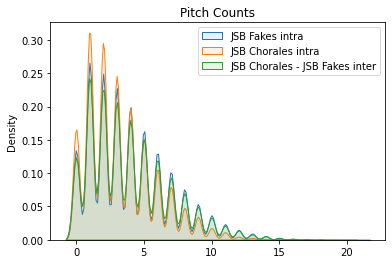}\\
    \end{minipage}%
    \begin{minipage}[b]{0.2\textwidth}
        \centering
        \includegraphics[width=\columnwidth]{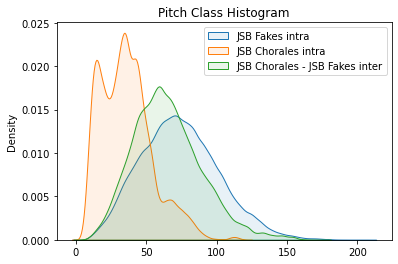}\\
    \end{minipage}%
    \begin{minipage}[b]{0.2\textwidth}
        \centering
        \includegraphics[width=\columnwidth]{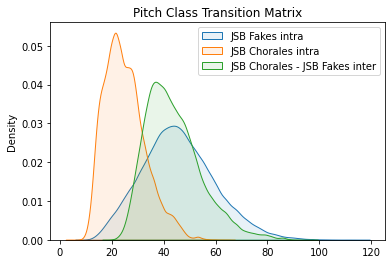}\\
    \end{minipage}%
    \begin{minipage}[b]{0.2\textwidth}
        \centering
        \includegraphics[width=\columnwidth]{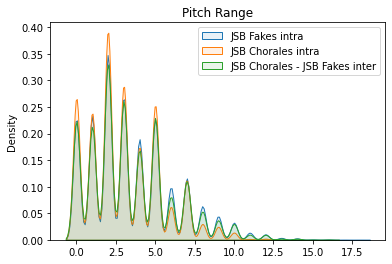}\\
    \end{minipage}%
    \begin{minipage}[b]{0.2\textwidth}
        \centering
        \includegraphics[width=\columnwidth]{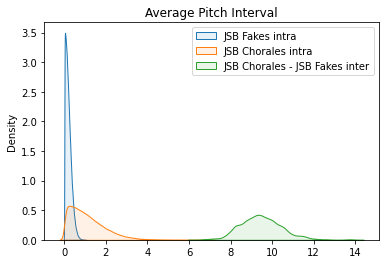}\\
    \end{minipage}%
    
    \centering
    \begin{minipage}[b]{0.2\textwidth}
        \centering
        \includegraphics[width=\columnwidth]{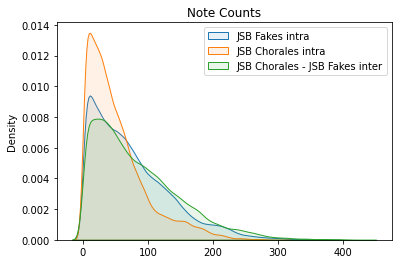}\\
    \end{minipage}%
    \begin{minipage}[b]{0.2\textwidth}
        \centering
        \includegraphics[width=\columnwidth]{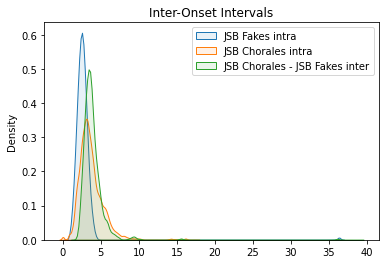}\\
    \end{minipage}%
    \begin{minipage}[b]{0.2\textwidth}
        \centering
        \includegraphics[width=\columnwidth]{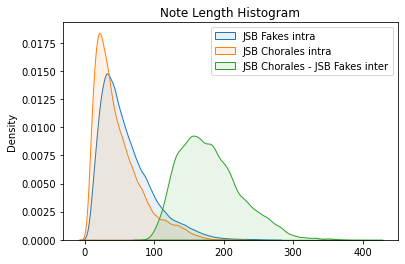}\\
    \end{minipage}%
    \begin{minipage}[b]{0.2\textwidth}
        \centering
        \includegraphics[width=\columnwidth]{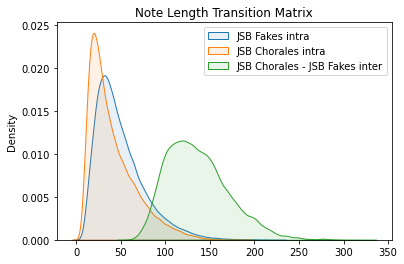}\\
    \end{minipage}%
    
    \caption{An example of objective evaluation of the \textit{intra-set} PDFs of the JSB Fakes and the JSB Chorales datasets. From left to right, up and down we show: a) Pitch Counts (PC) b) Pitch class Histogram (PCH) c) Pitch Class Transition Matrix (PCTM) d) Pitch Range (PR) e) Average Pitch Interval (PI) f) Note Counts (NC) g) Inter Onset Interval (IOI) h) Note Length Histogram (NLH) i) Note Length Transition Matrix (NLTM) PDFs.}
    \label{fig:eval}
\end{figure*}

\paragraph{Eval} We can evaluate music generation from a subjective perspective \cite{subjective} or an objective method. This submodule corresponds to the implementation of Yang and Lerch's evaluation method \cite{yang2020evaluation} for the objective evaluation of music generation. With it, we can cross-validate 2 or more data sets (trained and generated). To do so, we first calculate the Euclidean distances of the data sets from each other (inter-set) and from ecah other (intra-set), and construct the histograms of the calculated distances. We extract the probability density functions (PDF) from the histograms so that we can calculate the overlap area (OA) and the Kullback Leibler divergence (KLD). We can do this with each feature that we can calculate with the \texttt{features} submodule which are: \textit{Pitch count} (PC), \textit{Pitch class histogram} (PCH), \textit{Pitch class transition matrix} (PCTM), \textit{Pitch range} (PR) and \textit{Average pitch interval} (PI). The implemented rhythm-based features are: \textit{Note count} (NC) or \textit{Note Density}, \textit{Average inter-onset-interval} (IOI), \textit{Note length histogram} (NLH), \textit{Note length transition matrix} (NLTM).
An example of the PDFs extracted from 2 different datasets and features are shown in Fig. \ref{fig:eval}.

\subsubsection{Visualization} \label{sec:visual}
\begin{figure}
     \centering
     \includegraphics[width=\columnwidth]{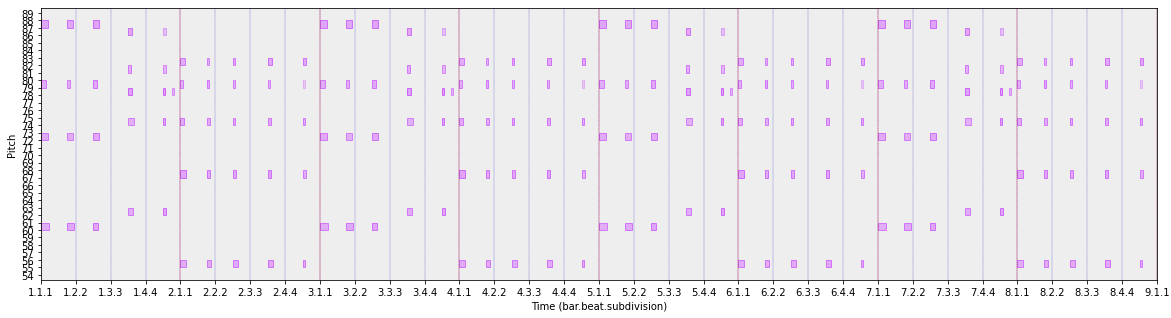}
     \caption{Pianoroll with subdivision's grid.}
     \label{fig:pianroll}
 \end{figure}

We can use the \texttt{musaiz} library also to plot symbolic music data as pianorolls. We provide two classes that allow us to generate and save the plots in \texttt{.png} (with \texttt{matplotlib}) and \texttt{.html} (with \texttt{plotly}) format. We set the x-axis (or time axis) with the bars and subdivisions that the user can set. This is useful to visualize the data as it is done in DAWs. We show an example of a pianoroll with the grid of subdivisions on the x-axis in Fig. \ref{fig:pianroll}.
 
\subsubsection{Exports} \label{sec:exports}
The latest functionality of the \texttt{musicaiz} software is the export function. The software provides a JSON representation of its objects that allows sending the music information via REST APIs, which makes the software useful for industrial applications as well.

\section{Illustrative examples}

\begin{lstlisting}[caption={Musicaiz loading, evaluation and visualization examples}, captionpos=b, language=iPython]
from musicaiz.loaders import Musa
from musicaiz.plotters import Pianoroll
from musicaiz import eval

# -------------Importing files----------------
# Read MIDI file
file = "mozart.mid"
midi = Musa(file, structure="bars", quantize=False)

# -------------Evaluation---------------------
# Extract features from MIDIs datasets and cross-validate
midis_path_1 = "midis_path_1/"
midis_path_1 = "midis_path_2/"
measures_1 = eval.get_all_dataset_measures(midis_path_1)
measures_2 = eval.get_all_dataset_measures(midis_path_2)
# Get the intra-set distances of dataset 1
intra_set_dists_1 = eval.euclidean_distance(measures_1)
# Get the intra-set distances
inter_set_dists = eval.euclidean_distance(measures_1, measures_2)

# -------------Plotting files----------------
# Plot the pianoroll of the 1st instrument
plot = Pianoroll()
plot.plot_instrument(
    total_bars=8,
    track=midi.instrument[0].notes,
    subdivision="quarter",
    print_measure_data=False,
    show_bar_labels=True,
)

\end{lstlisting}

\section{Impact}
Software is one of the foundations of AI research and is becoming essential to the scientific process. In MIR, researchers and engineers spend a lot of time building the processing or evaluation tools they use in their work, and sometimes these tools are often not published. By providing a general framework that not only follows best practices for scientific computing, but is also made following the principles of music, it is easier to build new models and reproduce the results, increasing the quality of scientific research. Therefore, \texttt{musicaiz} is built on top of well-established quality packages such as \texttt{prettymidi} and \texttt{mido} for handling symbolic musical information.

\texttt{MusicAIz} is tested with \texttt{pytest} and it is part of AIBeatz's\footnote{\url{https://app.aibeatz.com/}, last access July 2022} code base, a company that makes music beats with AI.

\section{Conclusions}
We have presented a new open source object-oriented library for symbolic music data that provides all the steps to generate and evaluate music with AI.
We hope that this library will be a useful tool for the community and will help to build music generation systems with (or without) AI approaches. Since it has been released as free and open source software, \texttt{musicaiz} can be easily updated to add other symbolic music formats such as MusicXML or adding more features for evaluating music generation such as structure pattern detection with self-similarity matrices \cite{foote1999visualizing} \cite{boundaries2021hernandezolivan}. Apart from that, further extensions can be made by adding new token representations that may be needed by future deep learning models or implementing new or other algorithms for feature extraction, either by us as original authors or by any other researcher or developer interested in it.

\section{Acknowledgments}
The authors would like to thank Ignacio Zay for his help in the implementation of the quantization algorithm.

This research has been partially supported by the Spanish Science, Innovation and University Ministry by the RTI2018-096986-B-C31 contract and the Aragonese Government by the AffectiveLab-T60-20R project.

\bibliographystyle{elsarticle-num} 
\bibliography{references}






\section*{Current code version}
\label{}

\begin{table}[!h]
\begin{tabular}{|l|p{6.5cm}|p{6.5cm}|}
\hline
\textbf{Nr.} & \textbf{Code metadata description} & \textbf{Please fill in this column} \\
\hline
C1 & Current code version & 0.0.2 \\
\hline
C2 & Permanent link to code/repository used for this code version & \url{https://github.com/carlosholivan/musicaiz} \\
\hline
C3  & Permanent link to Reproducible Capsule & \url{https://codeocean.com/capsule/3873536/tree} \\
\hline
C4 & Legal Code License   & GNU Affero General Public License\\
\hline
C5 & Code versioning system used & git\\
\hline
C6 & Software code languages, tools, and services used & python \\
\hline
C7 & Compilation requirements, operating environments \& dependencies & \url{https://github.com/carlosholivan/musicaiz/blob/master/requirements.txt}\\
\hline
C8 & If available Link to developer documentation/manual & \url{http://carlosholivan.github.io/musicaiz} \\
\hline
C9 & Support email for questions & carloshero@unizar.es\\
\hline
\end{tabular}
\caption{Code metadata (mandatory)}
\label{} 
\end{table}

\end{document}